\documentstyle[12pt,epsf]{article}
\begin{document}

\begin{center}
{\bf
SCALING PROPERTIES OF HADRON PRODUCTION \\[0.1cm]
IN $\pi^--p$ AND  $\pi^--A$ COLLISIONS AT HIGH-$P_T$ }


\vskip 5mm
G.P. \v {S}koro $^{1,*}$,
 M.V. Tokarev $^{2, \natural}$,
Yu.A. Panebratsev $^{2, \heartsuit}$
and  I.Zborovsk\'{y} $^{3, \clubsuit}$\footnote{The work is partially
supported by the  grant No. 020475 of the Czech Ministry of Education,
Youth and  Physical  Training.}

\vskip 5mm

{\small

(1) { \it Institute of Nuclear Sciences "Vin\v {c}a",\\
Faculty of Physics, University of Belgrade,\\
Belgrade, Yugoslavia}

\vskip 0.2cm

(2) {\it Laboratory of High Energies,\\
Joint Institute for Nuclear Research,\\
141980, Dubna, Moscow region, Russia}

\vskip 0.2cm

(3)  {\it Nuclear Physics Institute,\\
Academy of Sciences of the Czech Republic, \\
\v {R}e\v {z}, Czech Republic}

$^{*}$  E-mail: {\it {goran@rudjer.ff.bg.ac.yu}}

$^{\natural}$ {\it  E-mail: {tokarev@sunhe.jinr.ru}}

$^{\heartsuit}$ {\it  E-mail: {panebrat@sunhe.jinr.ru}}

$^{\clubsuit}$ {\it E-mail: {zborovsky@ujf.cas.cz}}

}
\end{center}

\vskip 5mm

\begin{center}
\begin{minipage}{150mm}
\centerline{\bf Abstract}
  Scaling features of  particles
  produced in $\pi^--p$ and $\pi^--A$ collisions over a high-$p_T$
  range at high energies are studied. The general concept of
  $z$-scaling is applied  for the analysis of $\pi^--p$ and
  $\pi^--A$ experimental data on the $Ed^3\sigma/dq^3$ inclusive
  cross section. The scaling function $\psi(z)$ and scaling variable
  $z$ are constructed and the anomalous dimension $\delta_{\pi}$ is found.
  The $A$-dependence of particle production
  in pion-nucleus collisions is studied.
The predictions
 of the inclusive cross section of the $\pi^{0}$-mesons produced
  in $\pi^--A$ collisions in the central rapidity range at
  high energies have been made.
\end{minipage}
\end{center}




{\section{Introduction}}

  Particle production with a high transverse momentum is traditionally
  connected with fundamental phenomena of
  elementary constituent interactions. The hypothesis of parton-hadron
  duality \cite{DPM} states, in particular,
  that the features of high-$p_T$ hadron spectra reflect the features
  of hard parton-parton interactions. It
  means that partons retain information about the collision during
  particle formation. Therefore, the features
  of single inclusive particle spectra in hadron-hadron and
  hadron-nucleus collisions of particle having a
  different flavour content are of interest to search for unusual
  properties of particle itself and its
  formation. Such features could be very useful to search for
  complementary signatures of unusual phenomena such
  as the phase transition of nuclear matter, new type of particle
  interactions and quark compositeness.

  One of the methods to study the properties of nuclear matter is
  to search for the violation of known scaling laws established
  in elementary collisions such as
  the Bjorken  and  Feynman  scaling laws.

  In this paper we study scaling features of hadron production
  in $\pi^--p$  and $\pi^--A$ collisions
  over a high $p_{T}$ range.
 Experimental data on
  cross section \cite{fris83,donal76,marzo87,turch93}
  and \cite{alver93,E706}   are used for the analysis.

 The $z$-scaling  was proposed in \cite{Z96} to
  describe the features of charged hadron production in $p-p$ and
  $\bar p-p$ collisions.
  New presentation ($z$-presentation) of experimental data can be obtained
  using the experimental observables,
  the inclusive cross section $Ed^3\sigma/dq^3$, the multiplicity
  density of charged particles
  $dN/d{\eta}|_{\eta=0}=\rho(s)$  and kinematical quantities of the
  reaction (collision energy, momenta and masses of initial  and produced
  particles).
  As shown in \cite{Z96,Z99,Z97,Z01} the data $z$-presentation
  reveals the symmetry properties,
  the  independence on  center-of-mass energy ${\sqrt s}$ and the angle
  of produced particle ${\theta}$ over a wide kinematical range.
  The energy dependence  of the $\pi^--p$ experimental data
  is used to find the anomalous dimension $\delta_{\pi}$.
  The scaling function $\psi(z)$ describes the
  probability density to form a particle with
  formation length $z$. The scaling variable $z$ reveals the property
  of fractal measure $z = z_0
  \epsilon^{-\delta}$, where $\epsilon $ is the scale resolution, and
  has a relevance to the geometry of
  space-time \cite{Zb,Z00}. It was shown \cite{Z01} that the A-dependence
  of high-$p_T$ hadron production in the
  framework of $z$-presentation is described by the function $\alpha (A)$
  depending on the single parameter, the  atomic weight $A$.

  The existence of the scaling  and its properties
  is assumed to reflect the fundamental features of
  particle structure, constituent interaction and particle
  production such as self-similarity, locality,
  fractality and scale-relativity.

\vskip 0.5cm
  {\section{ Z-scaling}}

 The idea of $z$-scaling is based on the assumption \cite{Stavinsky}
 that the  gross features of the inclusive particle
  distributions for reaction  (\ref{eq:r1}) at high energies
  \begin{equation}
  P_{1}+P_{2} \rightarrow q + X.
  \label{eq:r1}
  \end{equation}
can be described  in  terms of the corresponding kinematic
characteristics of the exclusive subprocess written in the symbolic
form
\begin{equation}
(x_{1}M_{1}) + (x_{2}M_{2}) \rightarrow m_{1} +
(x_{1}M_{1}+x_{2}M_{2} + m_{2}).
\label{eq:r2}
\end{equation}
The scale-invariant  fractions
$x_{1}$ and $x_{2}$ of the incoming four-momenta of colliding
objects are expressed via  momenta and masses of incident and produced
particles ($P_{1}, P_{2}, q$ and $M_{1}, M_{2}, m_1$) and determine
a minimum energy, which
 is necessary for the production of the secondary particle with mass $m_1$
 and  four-momentum $q$.
  The parameter $m_{2}$ is introduced  to satisfy the  internal conservation
  laws (for isospin, baryon number, and strangeness).

  \vskip 0.5cm
  {\subsection{Fractions $x_1$ and $x_2$}}

  The elementary parton-parton collision is considered  as a binary
  sub-process which satisfies the condition
  \begin{equation}
  (x_{1}P_{1} + x_{2}P_{2} - q)^{2} = (x_{1}M_{1} + x_{2}M_{2} +
  m_{2})^{2}.
  \label{eq:r5}
  \end{equation}
  The equation reflects minimum recoil mass hypothesis in the
  elementary sub-process. To connect kinematic and structural
  characteristics of the interaction, the coefficient
  $\Omega$ is introduced. It is chosen in the form
  \begin{equation}
  \Omega(x_1,x_2) = m(1-x_{1})^{\delta_1}(1-x_{2})^{\delta_2},
  \label{eq:r8}
  \end{equation}
  where $m$ is a mass constant and $\delta_1$ and $\delta_2$
  are factors relating to the fractal structure of
  the colliding objects \cite{Z99}\footnote{The anomalous dimensions
  are found to be  $\delta_{1,2} = \delta_h$ and
  $\delta_1 = \delta_h$, $\delta_2 = \delta_A = \delta_N \cdot A$
  for $h-h $ and $h-A$ collisions, respectively.}.
  The fractions $x_{1}$ and
  $x_{2}$  are determined  to maximize the value of $\Omega(x_1,x_2)$,
  simultaneously fulfilling condition
  $
  {d\Omega(x_1,x_2)/ dx_1}|_{x_2=x_2(x_1)} = 0.
  $
  Expressions for $x_{1}$ and $x_{2}$ as a
  function of the momenta and masses of
  the colliding and produced particles are given in \cite{Z99}.
  The variables
  $x_{1,2}$ are equal to unity along the phase space limit and
  cover the full phase space accessible at any
  energy.

\vskip 0.5cm
{\subsection{Scaling function $\psi(z)$ and variable $z$}}

 The scaling function $\psi$ expressed via the invariant differential
 cross section for the production of
 the inclusive particle $m_{1}$  is introduced as follows (see \cite{Z99})
 \begin{equation}
 \psi(z) = - \frac{\pi s_A}{\rho_A(s, \eta) \sigma_{inel}}J^{-1}
 E\frac{d\sigma}{dq^{3}}.
 \label{eq:r20}
 \end{equation}
 Here, $s_A \simeq s \cdot A$ and $ s $ are the center-of-mass energy
  squared of the corresponding $ h-A $
 and $ h-N $ systems and  $A$ is the atomic weight. The factor $J$ is
 a known function of kinematic variables
 \cite{Z99}. The expression (\ref{eq:r20}) relates the inclusive differential
  cross section and the average
 multiplicity density $\rho_A(s,\eta)$ to the scaling function $\psi(z)$.

The scaling function  is normalized as
\begin{equation}
\int_{z_{min}}^{\infty} \psi(z) dz = 1.
\label{eq:b6}
\end{equation}
The equation allow us to give the physical meaning
of the scaling function $\psi$ as a probability density to form
a particle  with a corresponding value of the variable $z$.

  The variable $z$ as argued in \cite{Z99}  can be interpreted
  as a particle formation length. It is chosen  in the form
\begin{equation}
z = \frac{ \sqrt{ {\hat s}_{\bot} }} {\Omega \cdot \rho_A(s) },
\label{eq:r28}
\end{equation}
where ${\hat{s}}^{1/2}_{\bot}$
is the transverse kinetic energy
of subprocess (\ref{eq:r2});
$\Omega$ is the measure given by (\ref{eq:r8}) and
$\rho_A(s) = \rho_A(s, \eta=0)$.
 We would like to note that the form of $z$ determines its variation range.
 The boundaries of the range are 0
 and $\infty$, as defined by (\ref{eq:r28})  and (\ref{eq:r8}).
 These values are scale independent and
 kinematically accessible at any energy.

  \vskip 0.5cm
  {\subsection{Fractality and scale-relativity}}

  Fractality in particle and nuclear physics concerns the internal
  structure of particles and their interactions. It is manifested
  by their self-similarity on any scale.
  This general principle is described by power law dependencies of
  the corresponding quantities \cite{Nottale,Z99}.

 The equation (\ref{eq:r5}) written in the form
  $x_1 x_2 - x_1 \lambda_2 - x_2 \lambda_1 = \lambda_0 $,
  does not change under the scale transformation
$
  \lambda_{1,2} \rightarrow \rho_{1,2} \cdot \lambda_{1,2},\ \ \
  x_{1,2} \rightarrow \rho_{1,2} \cdot x_{1,2}, \ \ \
  \lambda_0 \rightarrow  \rho_1 \cdot  \rho_2 \cdot  \lambda_0.$
   The transformation with the scale parameters
   $\rho_{1,2}$ allows us to consider the collisions of the complex
   objects in terms of suitable sub-processes of the
   interacting elementary constituents.
The coefficient $\Omega$, given by (\ref{eq:r8}),
 connects the kinematic and fractal characteristics of
 the interaction.
 The factors $\delta_1$ and $\delta_2$ are
 anomalous fractal dimensions of the colliding objects.
 The fractal structure itself is defined by the structure
  of the interacting constituents, which is not an elementary one either.
  In this scheme, high energy hadron-hadron, hadron-nucleus and
  nucleus-nucleus interactions are considered as interactions of fractals.


  The variable  $z$ written in the form
  $ z=z_0\cdot {\epsilon}^{-\delta}$ (where
   $z_0=\sqrt{{\hat s}_{\bot}}/\rho(s)$ and
 $\epsilon ^{-1} =[m(1-x_1)^{A_1}(1-x_2)^{A_2}]^{-1} $)
 reveals the properties  of a fractal measure
 and $\delta $ is the anomalous fractal dimension describing
 the intrinsic structure of the interaction constituents
 revealed at high energies. The nontrivial features of
 mechanism of particle formation is that the formation length $z$
 increases with resolution $\epsilon ^{-1}$.


  \vskip 0.5cm
 { \section{ $\pi^-p$,  $\pi^--$nucleus collisions and $z$-scaling}}

   Experimental data sets
   of cross sections for $\pi^{\pm,0}, K^{\pm}, \bar p$ hadrons
   produced  in $\pi^--p$ and $\pi^--A$ collisions at high transverse
   momentum $p_T$  are presented in
   \cite{fris83,donal76,marzo87,turch93} and \cite{alver93,E706},
   respectively.
   The measurements were made at
   pion momentum $p_{lab}=40, 200, 300~GeV/c$  over the
   range  $0.8<p_T<10.~GeV/c$. The nuclear  targets  $Be, Cu$
   and $W$ were used.

    We would like to note that all the cross section data
    demonstrate  the strong energy dependence of
 the cross section on transverse momentum,
 the tendency  that  difference between hadron yields increases
 with transverse momentum and energy $\sqrt s$ and
 the non-exponential behavior of the spectra
 at $p_{T}>1~GeV/c$.

  \vskip 0.5cm
{\subsection{$\pi^--p$ collisions}}

 In this section we study the properties of $z$-scaling
 for hadrons produced in $\pi^--p$ collisions.
  The charged particle density $\rho (s,\eta) $  in $\pi^--p$ collisions was simulated
 by PYTHIA \cite{PYTHIA} in the energy range $\sqrt s = 10-200~GeV$.
 The results of simulations show that the energy dependence of the
 density $\rho (s)$ for processes  $\pi^--p$ and $p-p$ is practically
 the same one (taking into account the errors)
 and can be parameterized in the form $\rho = a s^b$.
 The values of the parameters were found to be
 $a = 0.74\pm 0.12,\ b =0.105\pm 0.011$ and
 $a = 0.59\pm 0.08,\ b =0.126\pm 0.017$ for  $p-p$ and  $\pi^--p$,
 respectively. The PYTHIA results give the relation
 $\sigma_{\pi p} = 0.67 \sigma_{pp}$ expected from quark counting rule too.
 We do not have enough experimental data for $\rho(s) $ of
 $\pi^--p$ at high  $\sqrt s$ and $p_T$ and the available
 experimental data \cite{NA22} are not in disagreement with MC
 results. Therefore we use in our analysis of $z$-scaling
 in $\pi^--p$  collisions
 the experimentally measured dependence of the average charged particle
 multiplicity density  for $p-p$ collisions. As we will show later
 the replacement does not destroy the general properties of
 $z$-scaling  in $\pi^--p$ particle production.

 We verify the hypothesis of energy scaling for data
 $z$-presentation for hadron production
 in $\pi^--p$ collisions using the available experimental data.

 Figures 1(a)-2(a)  show the dependence of the cross section
 $Ed^3\sigma/dq^3$  of  $\pi^{+}$ and $K^{-}$-mesons produced
 in $\pi^--p$
 on transverse momentum $p_{T}$ at $p_{lab} = 40, 200 ,300~GeV$
 and the produced angle $\theta_{cm}$ near $90^0$.
 Note that the data cover the  wide
 transverse momentum range, $p_{T}=1-6~GeV/c$.


 Figures 1(b)-2(b) show $z$-presentation of the same data sets.
 Taking into account the experimental errors we can conclude that
 the scaling function $\psi(z)$ demonstrates an energy
 independence over a wide energy and transverse momentum
 range at $\theta_{cm}^{\pi N} \simeq 90^0$.
 The energy dependence  of the $\pi^--p$ experimental data
 is used to find the value of the anomalous dimension $\delta_{\pi}$.
 It is equal to 0.1.



 To analyze the angular dependence
 of the scaling function $\psi(z)$ of charged hadrons
 $\pi^{\pm}, K^{\pm}, \bar p$ produced
 in $\pi^--p$ collisions  we use the data set obtained
 at Protvino \cite{turch93}.
 The data set includes the results
 of measurements of the invariant cross section
 $Ed^3\sigma/dq^3$  at the pion incident momentum
 $p_{lab} = 40~GeV$ over the momentum and angular ranges of
 $p_{T}=1.05-3.75~GeV/c$ and $\theta_{cm}^{\pi N} = 49^0-93^0$.
 The obtained results show that experimental errors are large enough
 and more high accuracy data on the
 cross section are necessary to verify carefully the
 angular independence of the scaling function $\psi(z)$ of hadrons
 produced in $\pi^--p$ collisions  as a function
 of energy $\sqrt s $, transverse momentum $p_T$ and
 and $\theta_{cm}^{\pi N}$.

\vskip  0.5cm
{\subsection{$\pi^--A$ collisions} }

 In this section, we study the properties of $z$-scaling for
 hadron  production  in $\pi^--$nucleus
 collisions. The  experimental data sets \cite{fris83,alver93} and
 \cite{E706} are used in the analysis.

  According to the procedure of
  $z$-analysis of the $p-A$ experimental data, the function $\psi$
  is calculated for every nucleus using the normalization  factor
  $\sigma_{inel}^{pA}/\sigma_{inel}^{pp}$ \cite{Z01}
  in the expression for the inclusive cross section
  \footnote{The other normalization  factor,
  $\sigma_{inel}$, was used in \cite{Z99}.}.
  The factor $\sigma_{inel}^{pA}$ is the total inelastic
 cross section for $pA$ interactions.
 The $A$-dependence of the ratio $\sigma_{inel}^{pA}/\sigma_{inel}^{pp}$
 is taken from \cite{Carroll}.
  The relevant multiplicity
 densities of charged particles obtained  by the Monte Carlo
 simulation generator  HIJING \cite{Hij1} for different nuclei
 ($A=7-197$) are taken in the form  $\rho_A(s) {\simeq }0.67{\cdot}
 A^{0.18}{\cdot }s^{0.105} $ \cite{Z99}.
 In the present analysis  we use for the
 multiplicity density $\rho_A(s)$
 and the $A$-dependence of the ratio $\sigma_{inel}^{\pi A}/\sigma_{inel}^{\pi p}$
 of $\pi^--A$
 the results obtained for $p-A$ collisions. The possibility of such replacement
 will be argued by obtained results for data $z$-presentation.

 The symmetry
 transformations
$ z \rightarrow \alpha (A) \cdot z,
 \ \psi \rightarrow \alpha^{-1} (A) \cdot \psi $
 of the function  $\psi(z)$  and the argument $z$ are used to
 compare the functions $\psi$  for different nuclei.

 Figures 3(a),4(a) show the dependence of the inclusive cross
 section  for $\pi^{+}, K^{-}$
 produced in $\pi^--Be, Cu$, and $W$ collisions on the transverse
 momentum $p_{T}$ at $p_{lab} = 200$ and $300~GeV/c$.
 The incisive cross section data for $W$ nucleus demonstrate the
 energy dependence, which enhances as the transfers momentum of
 produced particle increases.

 Figure 5(a) presents cross sections for $\pi^0$-mesons produced
 on $Be$ and $Cu$ nuclear targets at $\sqrt s \simeq  31~GeV$.

 The dependencies of the scaling function $\psi$ on $z$ of the same
 experimental data are shown in Figures 3(b)-5(b).
 As seen from Figure 5(b) the asymptotic regime
 (the power law for the scaling function, $\psi(z)\sim
 z^{-\beta}$) is achieved over a high-$p_T$ range for $\pi^0$-meson
 production on nuclei $Be$ and $Cu$ at $\sqrt s \simeq 31~GeV$.
 The value of the slope parameter $\beta$
 is found to be $\simeq 9.37$  over a wide range of
 high transverse momentum ($3<q_{T}<7.5~GeV/c$).

 We use the properties of $z$-scaling to calculate the
 cross section of $\pi^0$-meson production in $\pi^-Be$ and
 $\pi^-Cu$ collisions at high energies. The results are shown in
 Figures 6(a,b).

\vskip  0.5cm
{\section{Conclusions}}

 The scaling features of $\pi^{+}, K^{-}$ hadrons
 produced in $\pi^--p$ and $\pi^--A$ collisions at high energies in
 terms of $z$-scaling are studied. The experimental data sets
 \cite{fris83,donal76,marzo87,turch93} and \cite{alver93,E706} on
 the inclusive cross sections are used in the analysis. The
 momentum of incident pion beam $p_{lab}$ changes from 40 to $515~GeV/c$
 over the high transverse momentum range $(p_{T}=0.2-10~GeV/c)$.

 The $z$-presentation  of experimental data is constructed
 and  the anomalous fractal dimension $\delta_{\pi}$ is found
 to be 0.1.
 The value is allowed us to reproduce the general properties of
 $z$-scaling established in $p-p$, $\bar p-p$ and $p-A$ collisions.


 The $A$-dependence of data $z$-presentation is studied
 and it is shown that the dependence is described by the
 function  $\alpha = \alpha(A)$. The fractal
 dimension of nuclei $\delta_A$ for $\pi^--A$ is found to be the
 same  $\delta_A= A\cdot \delta_N$ as for the
 hadrons produced in $p-A$ collisions.

 The asymptotic regime of the scaling function, $\psi(z)\sim
 z^{-\beta}$, is observed  and the asymptotic value of the slope
 parameter  $\beta $ is determined  to be  $\simeq 9.37$
 at $\sqrt s \simeq  31~GeV$ and over the range
 $3 <p_{T}< 7.5~GeV/c$.

 Using the properties of $z$-scaling, the dependence of the cross
 sections of $\pi^{0}$-mesons produced in $\pi^--Be$, $\pi^--Cu$
 and $\pi^--Au$ collisions on transverse momentum over the central
 rapidity range at high energy $\sqrt s = 60, 200$ and $500~GeV$ is
 predicted.


 Thus, the obtained results show that data $z$-presentation of
 hadrons  produced in $\pi^--p$ and
 $\pi^--A$ collisions demonstrates general properties of the
 particle formation mechanism such as self-similarity, locality,
 scale relativity and fractality.  As one can assume the properties
 reflect through the anomalous dimension $\delta_{\pi}$
 the features of elementary constituent substructure too.






%
%

{\small


\newpage
\begin{minipage}{4cm}

\end{minipage}

\vskip 4cm
\begin{center}
\hspace*{-2.5cm}
\parbox{5cm}{\epsfxsize=5.cm\epsfysize=5.cm\epsfbox[95 95 400 400]
{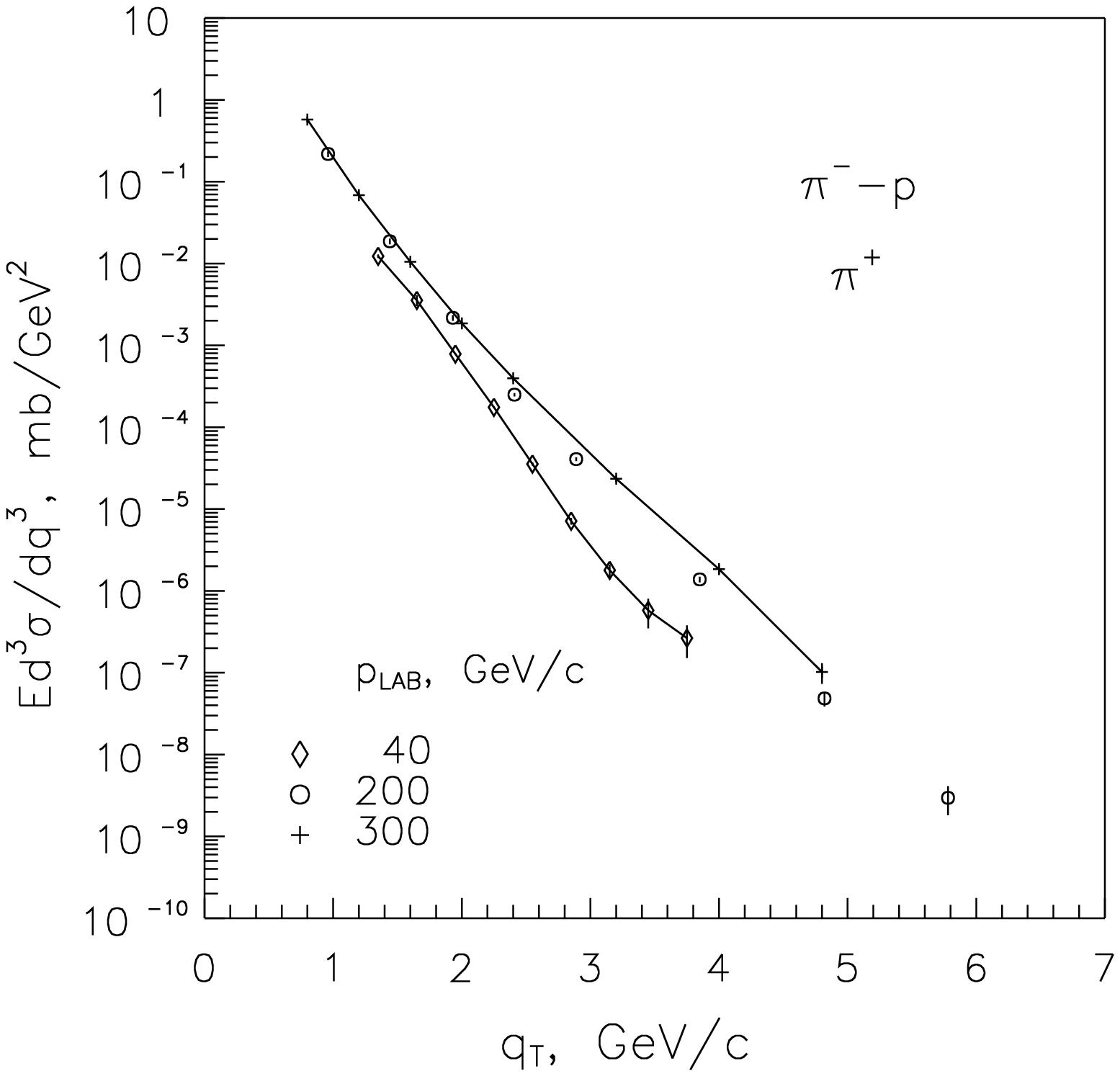}{}}
\hspace*{3cm}
\parbox{5cm}{\epsfxsize=5.cm\epsfysize=5.cm\epsfbox[95 95 400 400]
{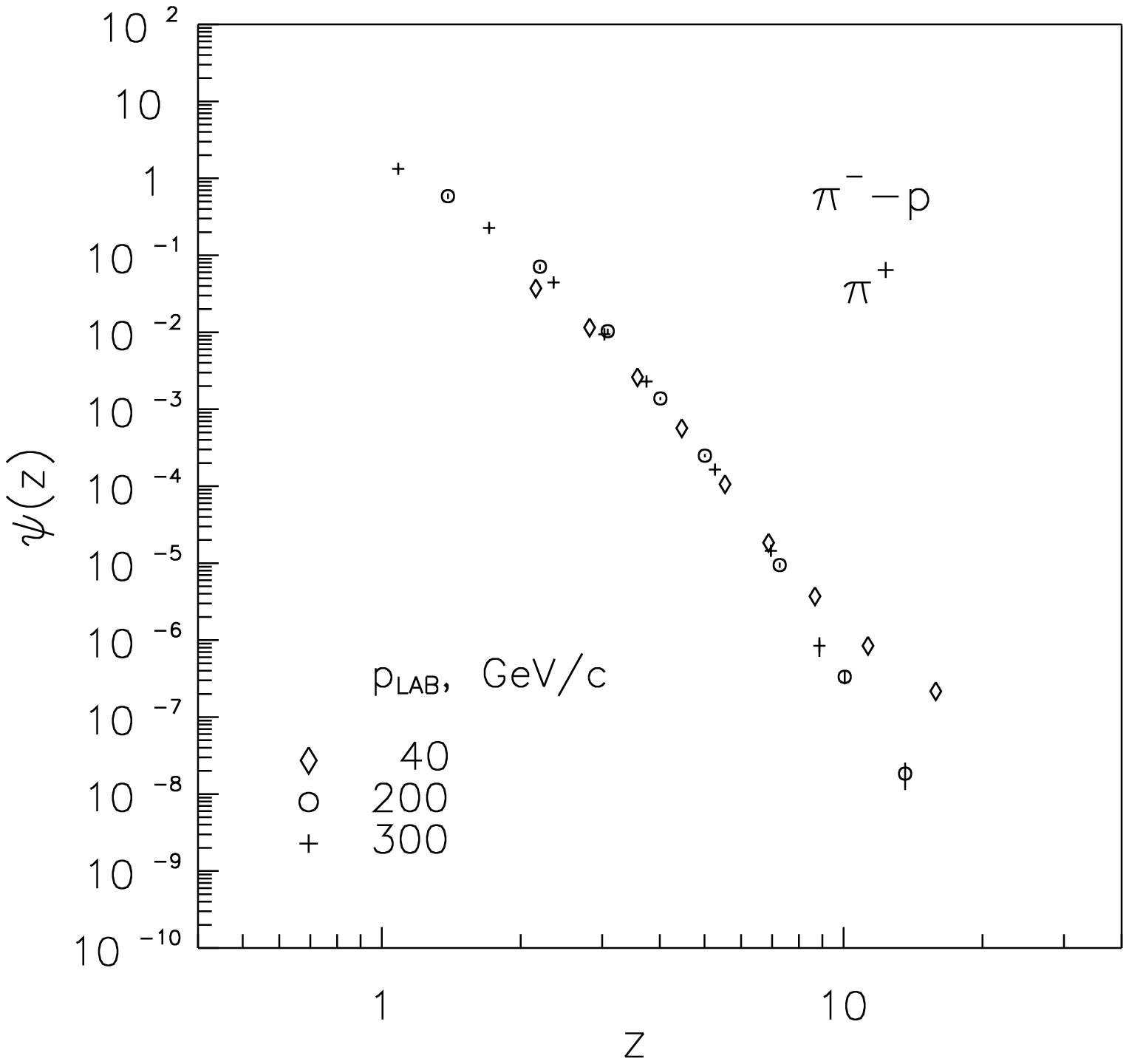}{}}
\vskip -0.5cm
\hspace*{0.cm} a) \hspace*{8.cm} b)\\[0.5cm]
\end{center}

 {\bf Figure 1.}
 (a) Dependence of  the
 inclusive cross section of $\pi^+$-meson  production
 on transverse momentum $q_{T}$ at $p_{lab} = 40, 200$ and $300~GeV/c$
 and $\theta_{cm}^{\pi p} \simeq 90^{0}$
 in  $\pi^--p$ collisions.
 Experimental data are taken from
 \cite{fris83,turch93}.
 (b) The corresponding scaling function $\psi(z)$.


\vskip 5cm

\begin{center}
\hspace*{-2.5cm}
\parbox{5cm}{\epsfxsize=5.cm\epsfysize=5.cm\epsfbox[95 95 400 400]
{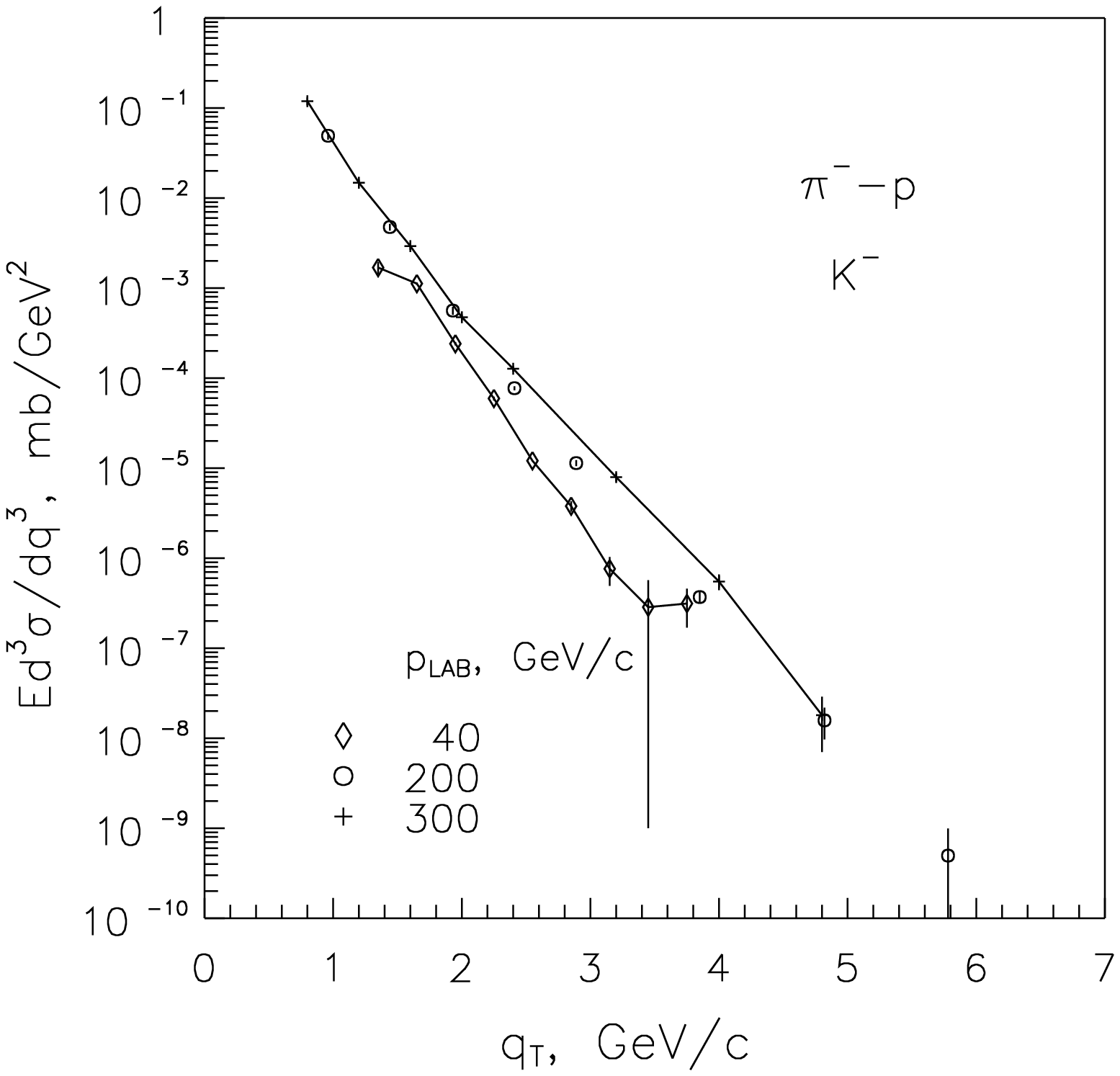}{}}
\hspace*{3cm}
\parbox{5cm}{\epsfxsize=5.cm\epsfysize=5.cm\epsfbox[95 95 400 400]
{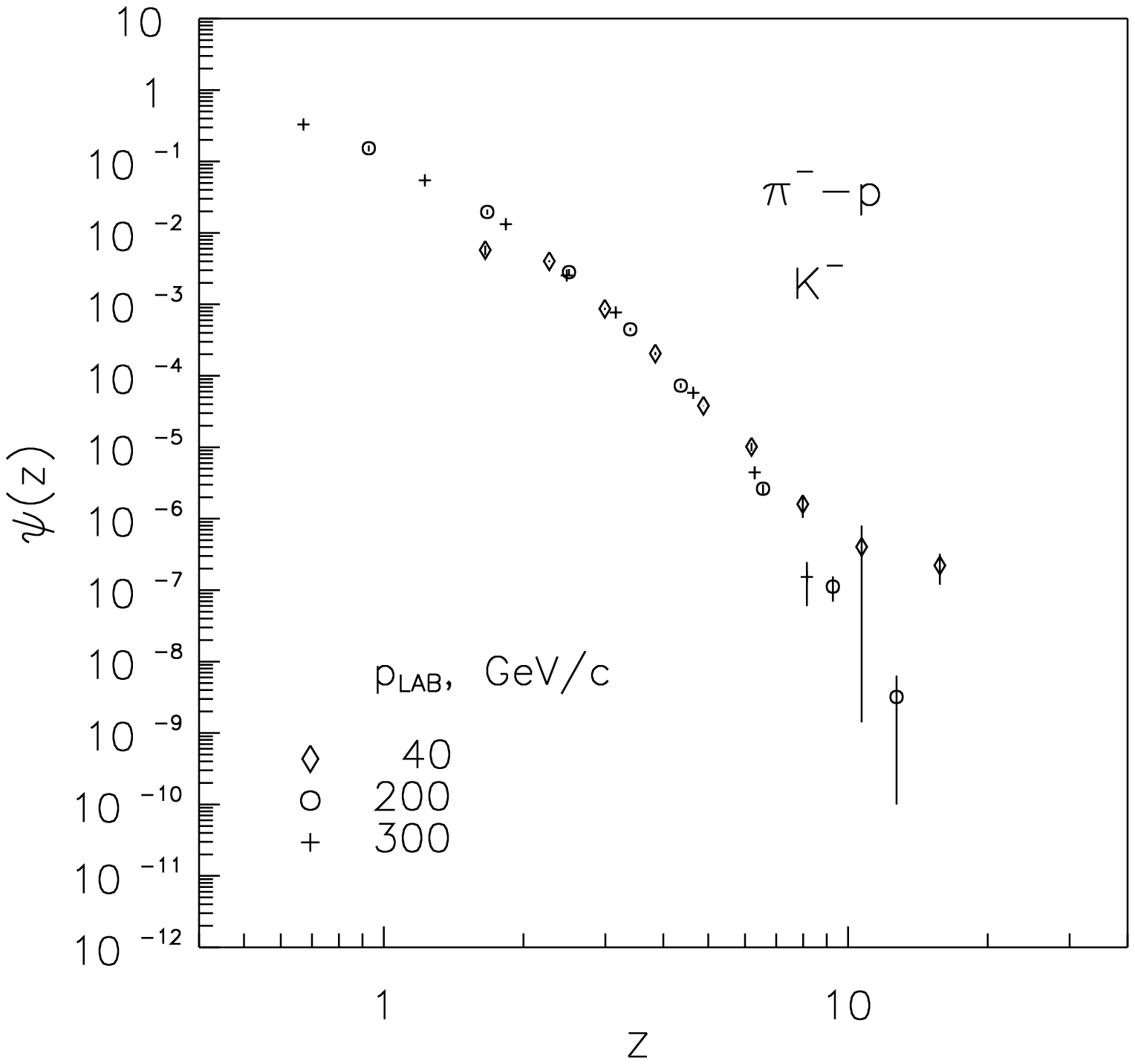}{}}
\vskip -1.cm
\hspace*{0.cm} a) \hspace*{8.cm} b)\\[0.5cm]
\end{center}

 {\bf Figure 2.}
 (a) Dependence of  the
 inclusive cross section of $K^-$-meson  production
 on transverse momentum $q_{T}$ at $p_{lab} = 40, 200$ and $300~GeV/c$
 and $\theta_{cm}^{\pi p} \simeq 90^{0}$
 in  $\pi^--p$ collisions.
 Experimental data are taken from
 \cite{fris83,turch93}.
 (b) The corresponding scaling function $\psi(z)$.


\newpage
\begin{minipage}{4cm}

\end{minipage}

\vskip 4cm
\begin{center}
\hspace*{-2.5cm}
\parbox{5cm}{\epsfxsize=5.cm\epsfysize=5.cm\epsfbox[95 95 400 400]
{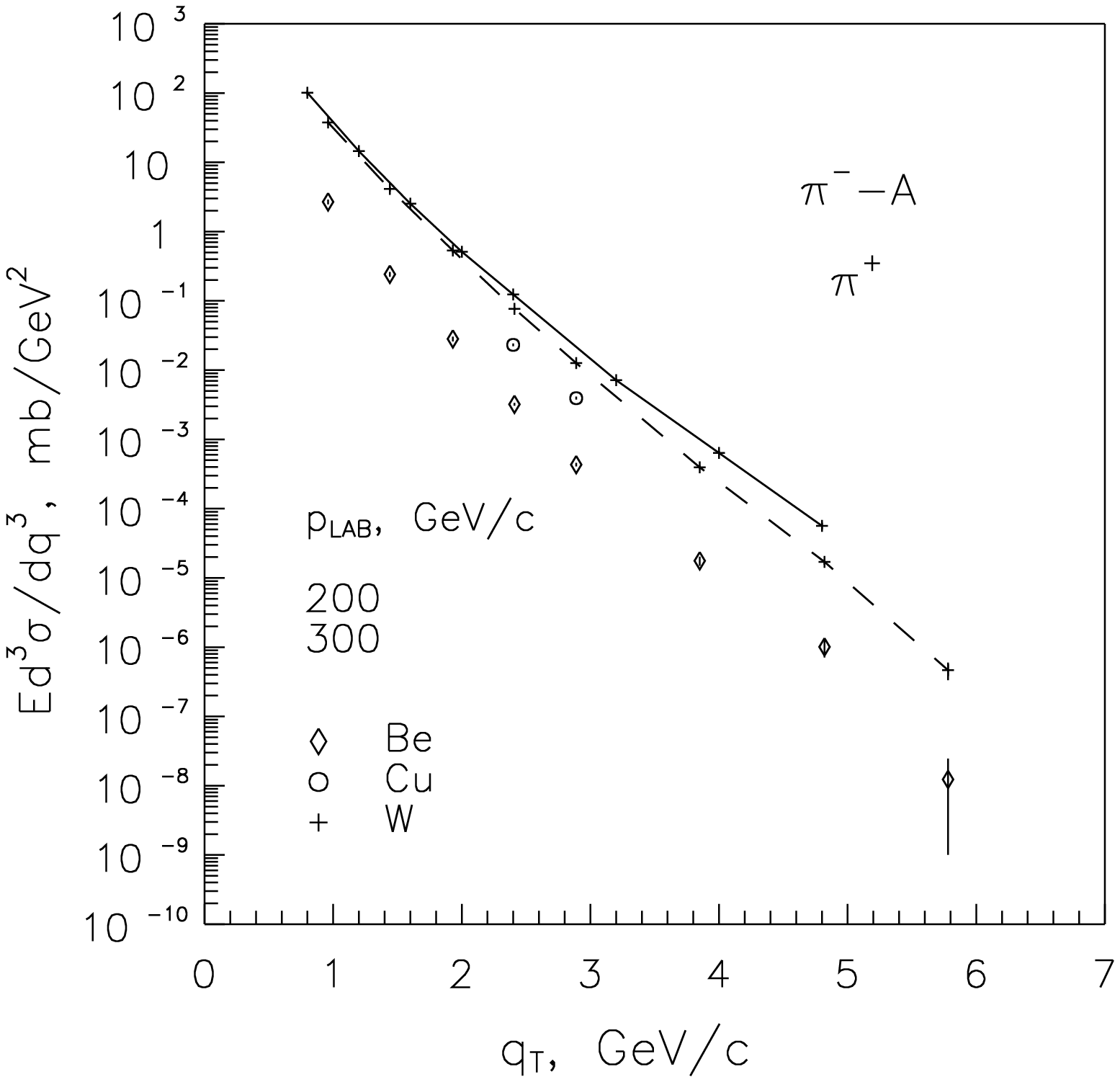}{}}
\hspace*{3cm}
\parbox{5cm}{\epsfxsize=5.cm\epsfysize=5.cm\epsfbox[95 95 400 400]
{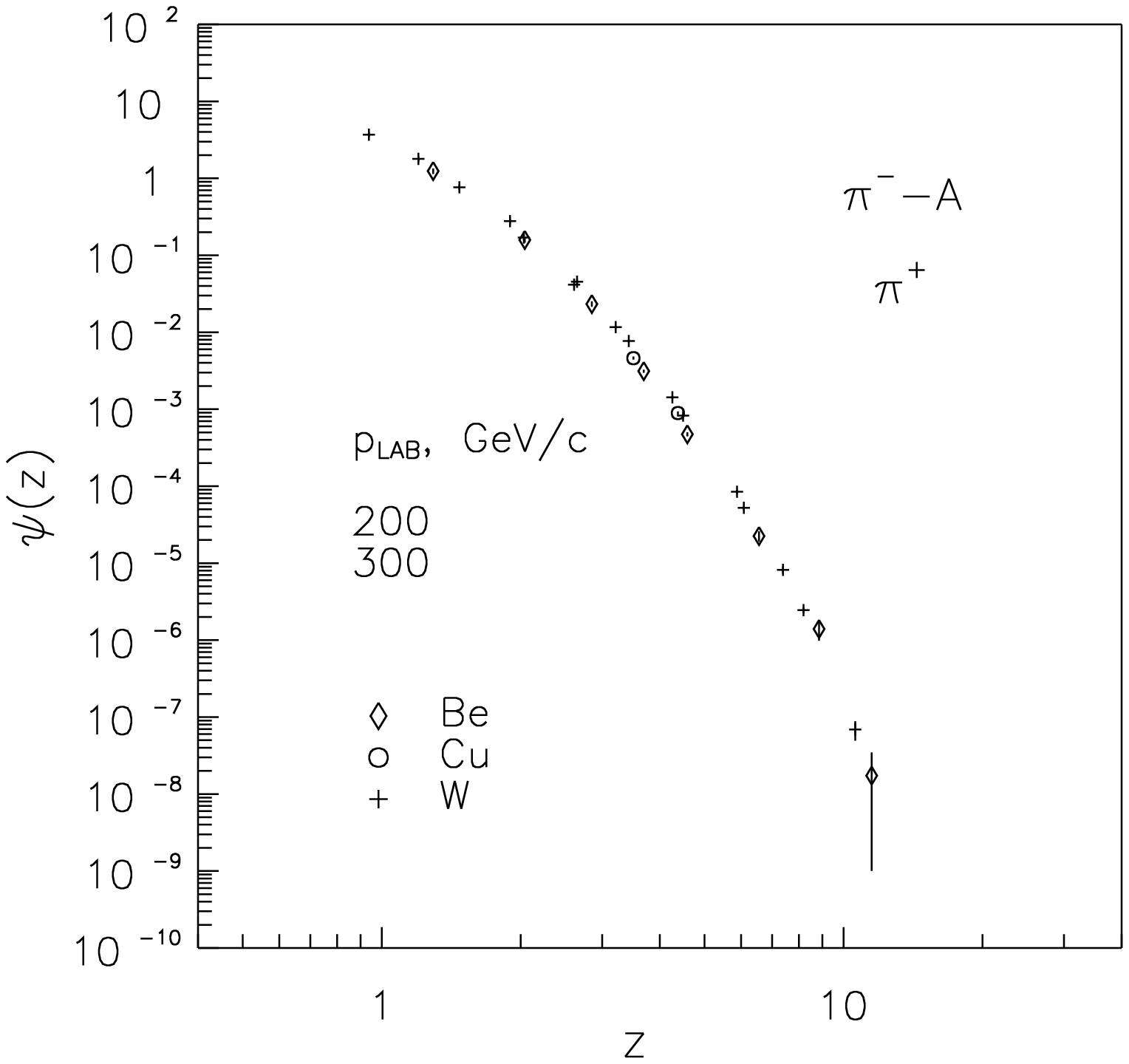}{}}
\vskip -0.5cm
\hspace*{0.cm} a) \hspace*{8.cm} b)\\[0.5cm]
\end{center}

{\bf Figure 3.}
(a) Dependence of  the
inclusive cross section of $\pi^+$-meson  production
on transverse momentum $q_{T}$
in  $\pi^--A$ collisions
at $p_{lab} = 200, 300~GeV/c$.
Experimental data are taken from
\cite{fris83}.
(b) The corresponding scaling function $\psi(z)$.


\vskip 5cm

\begin{center}
\hspace*{-2.5cm}
\parbox{5cm}{\epsfxsize=5.cm\epsfysize=5.cm\epsfbox[95 95 400 400]
{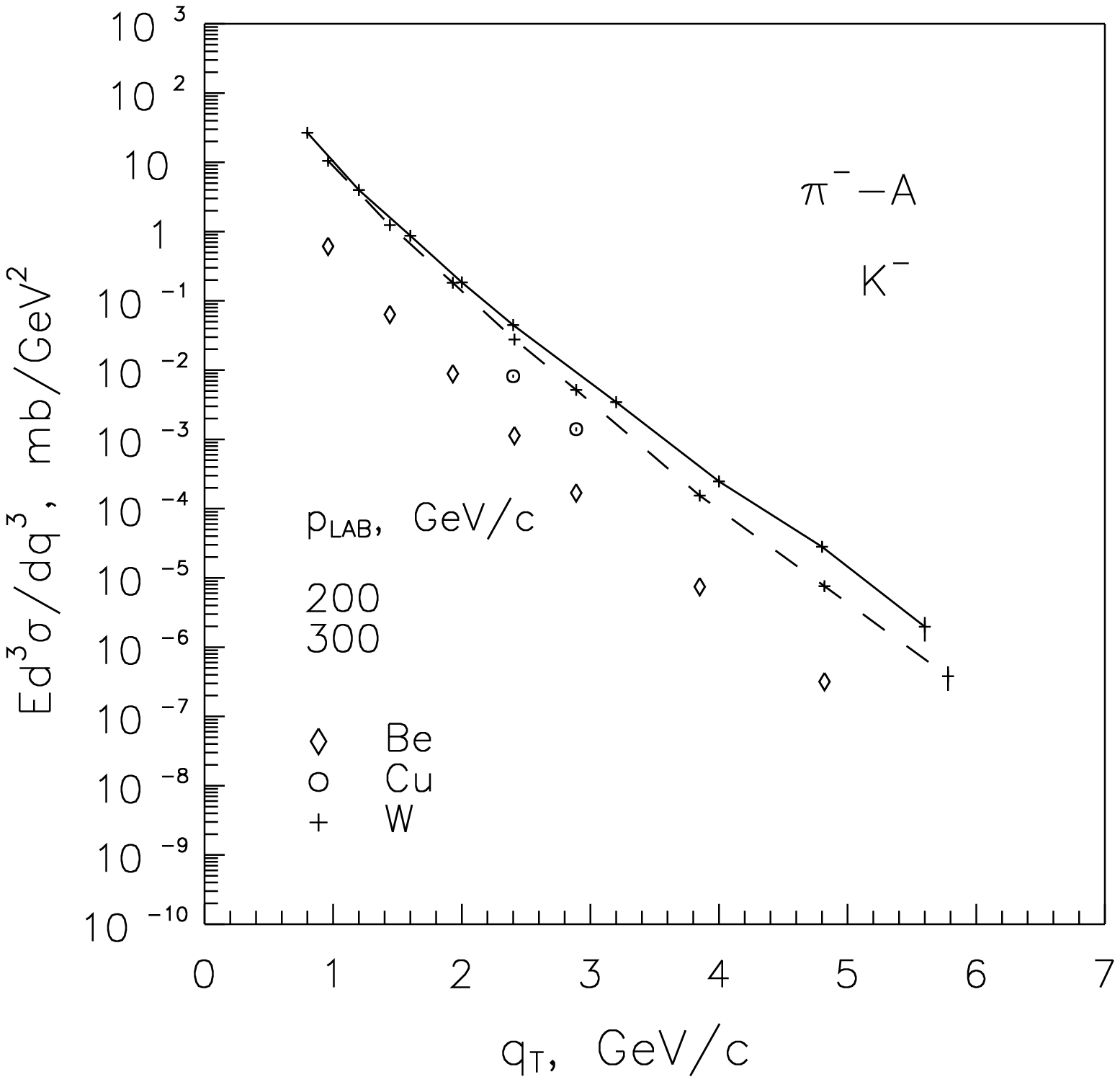}{}}
\hspace*{3cm}
\parbox{5cm}{\epsfxsize=5.cm\epsfysize=5.cm\epsfbox[95 95 400 400]
{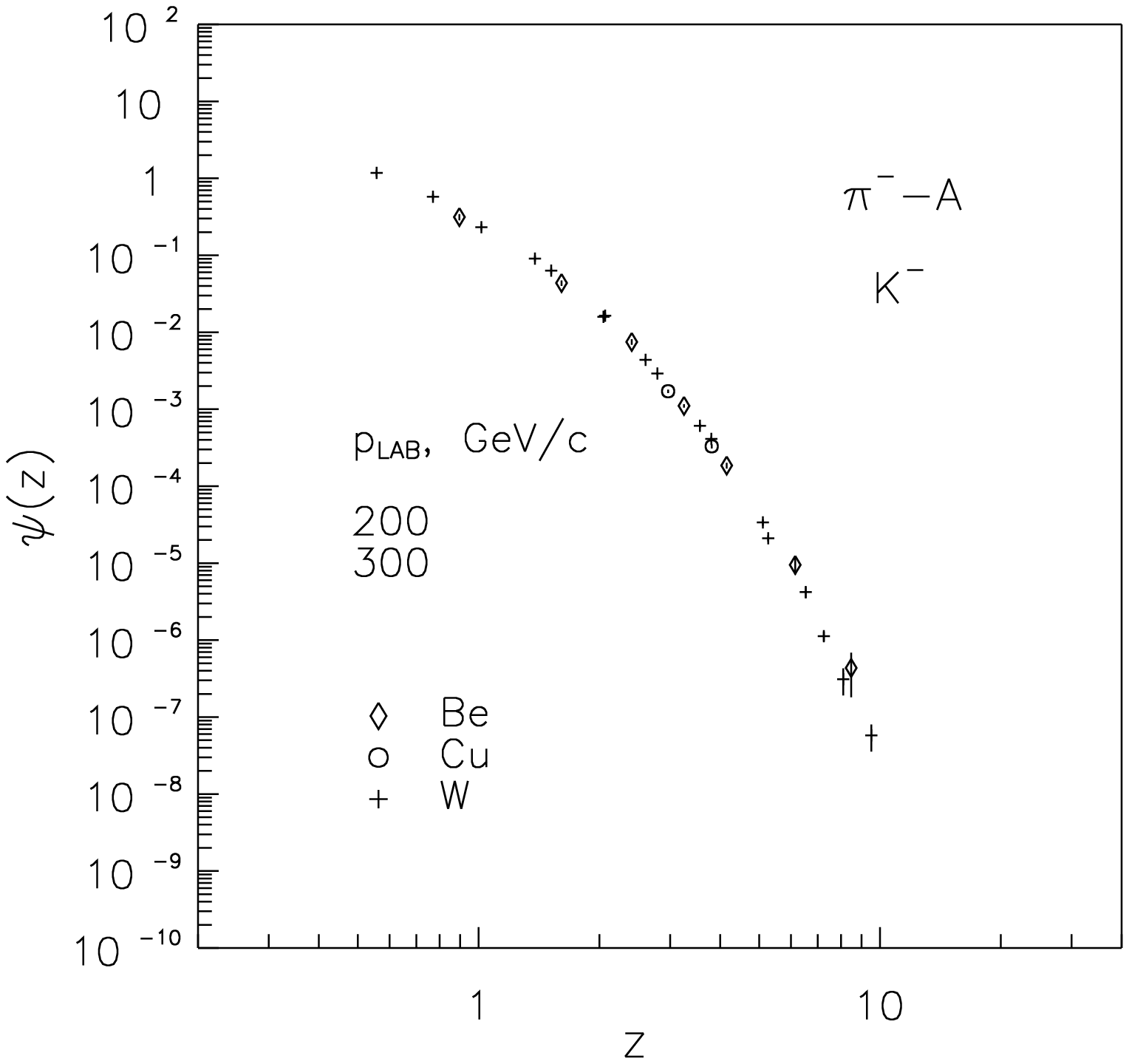}{}}
\vskip -1.cm
\hspace*{0.cm} a) \hspace*{8.cm} b)\\[0.5cm]
\end{center}

{\bf Figure 4.}
(a) Dependence of  the
inclusive cross section of $K^-$-meson  production
on transverse momentum $q_{T}$
in  $\pi^--A$ collisions
at $p_{lab} = 200, 300~GeV/c$.
Experimental data are taken from
\cite{fris83}.
(b) The corresponding scaling function $\psi(z)$.


\newpage
\begin{minipage}{4cm}

\end{minipage}

\vskip 4cm
\begin{center}
\hspace*{-2.5cm}
\parbox{5cm}{\epsfxsize=5.cm\epsfysize=5.cm\epsfbox[95 95 400 400]
{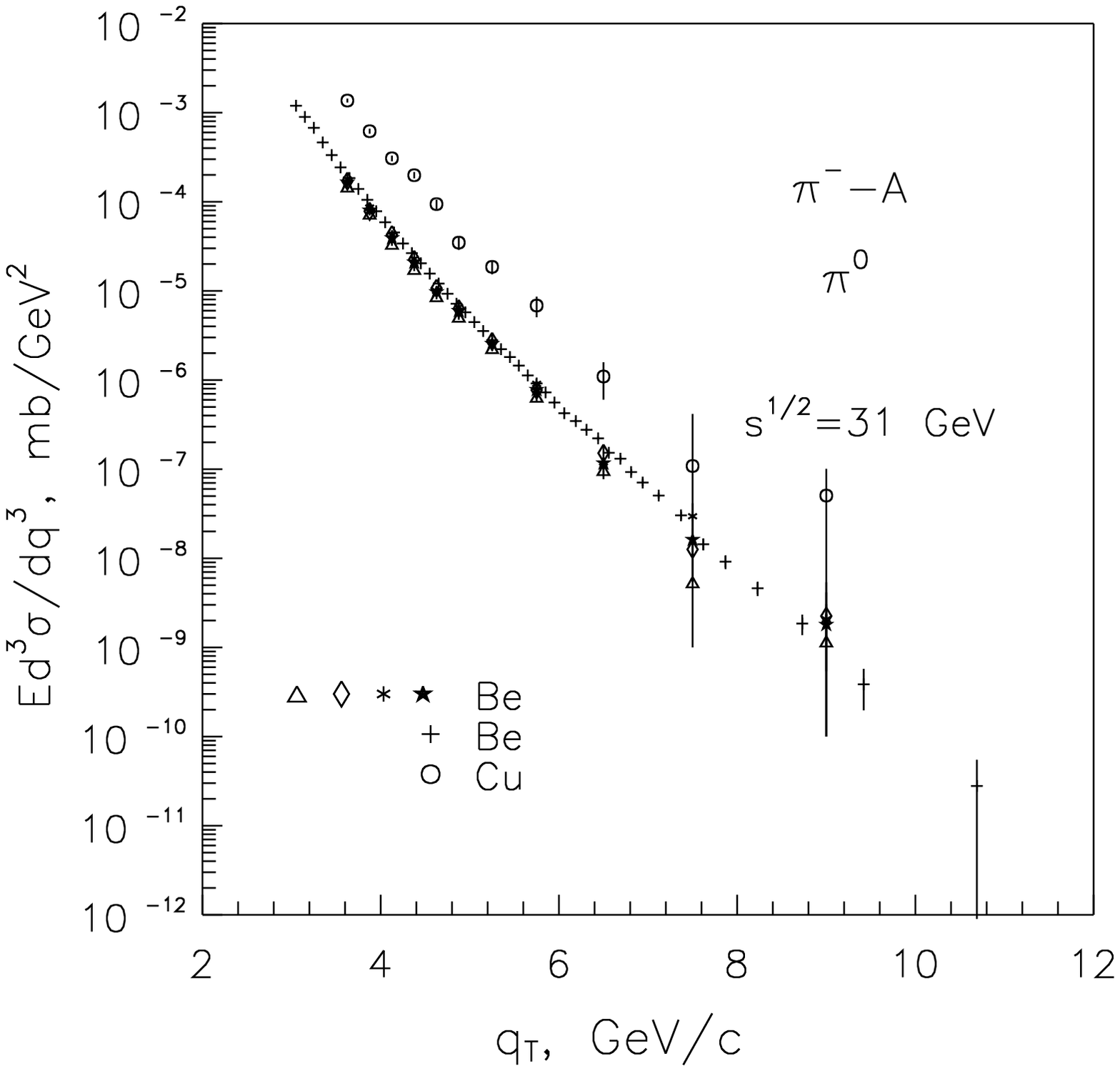}{}}
\hspace*{3cm}
\parbox{5cm}{\epsfxsize=5.cm\epsfysize=5.cm\epsfbox[95 95 400 400]
{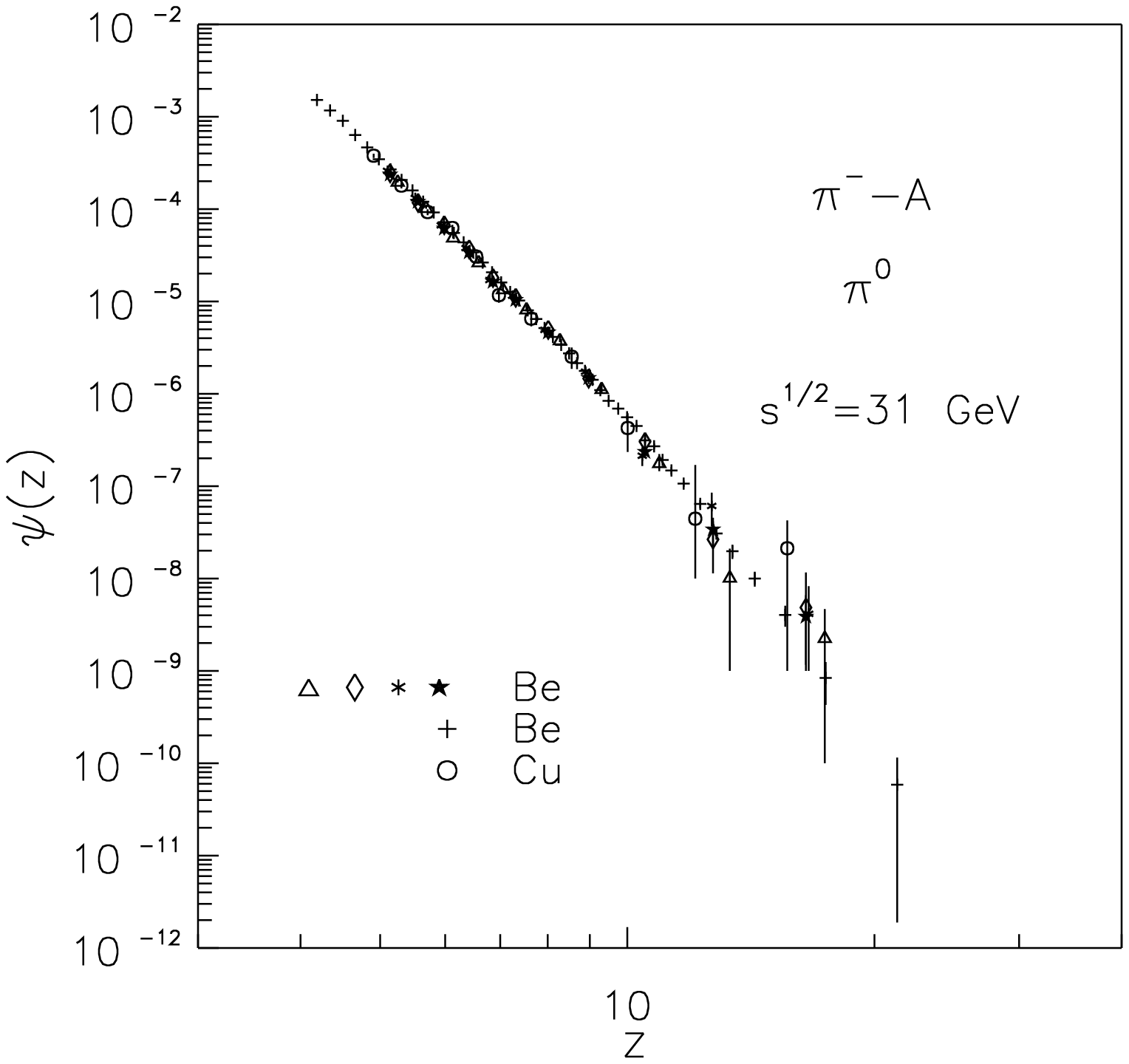}{}}
\vskip -0.5cm
\hspace*{0.cm} a) \hspace*{8.cm} b)\\[0.5cm]
\end{center}

{\bf Figure 5.}
(a) Dependence of  the
inclusive cross section of $\pi^0$-meson  production
on transverse momentum $q_{T}$
in  $\pi^--A$ collisions
at $\sqrt s =31~GeV$.
Experimental data are taken from
\cite{alver93,E706}.
(b) The corresponding scaling function $\psi(z)$.


\vskip 5cm

\begin{center}
\hspace*{-2.5cm}
\parbox{5cm}{\epsfxsize=5.cm\epsfysize=5.cm\epsfbox[95 95 400 400]
{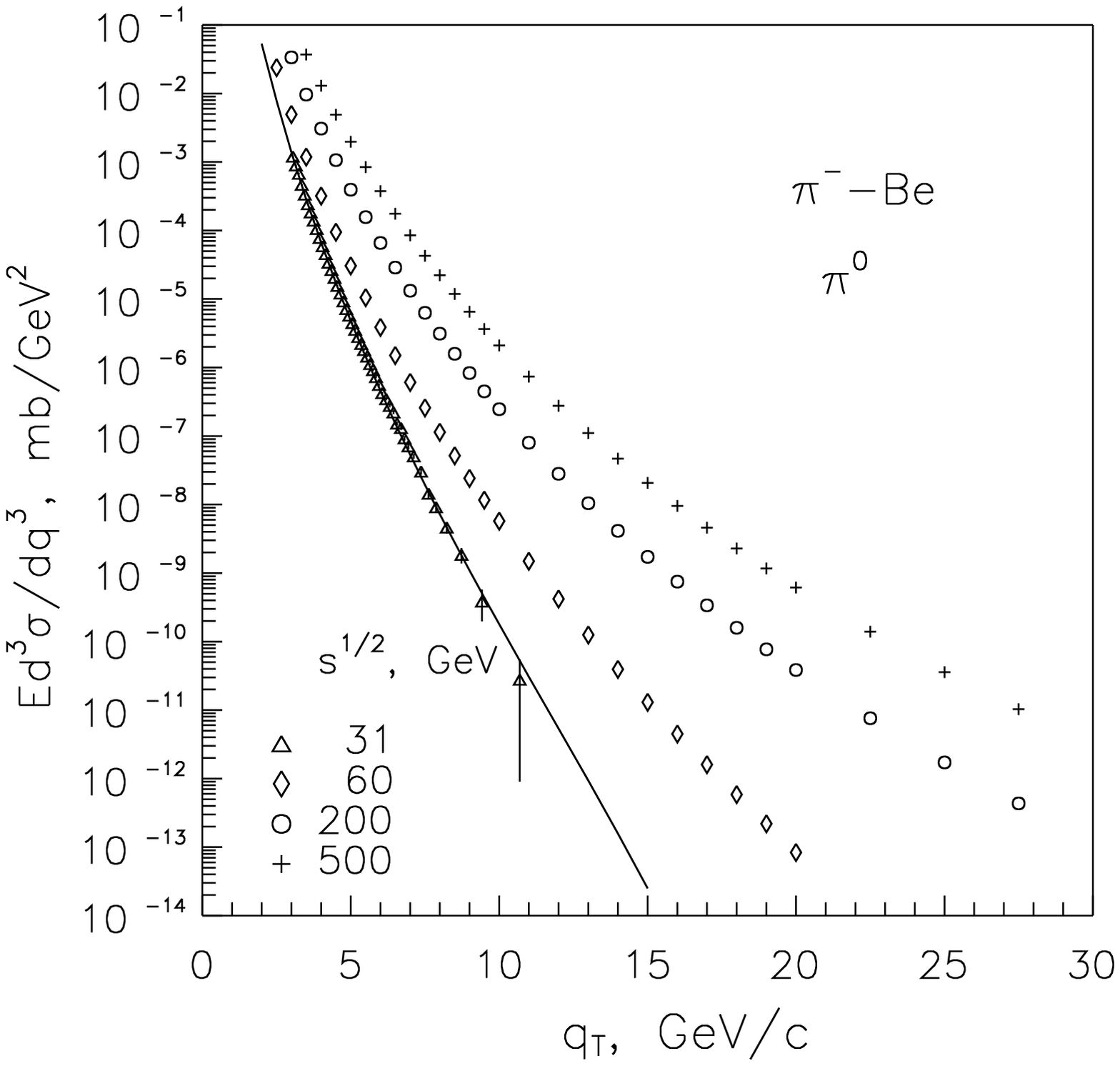}{}}
\hspace*{3cm}
\parbox{5cm}{\epsfxsize=5.cm\epsfysize=5.cm\epsfbox[95 95 400 400]
{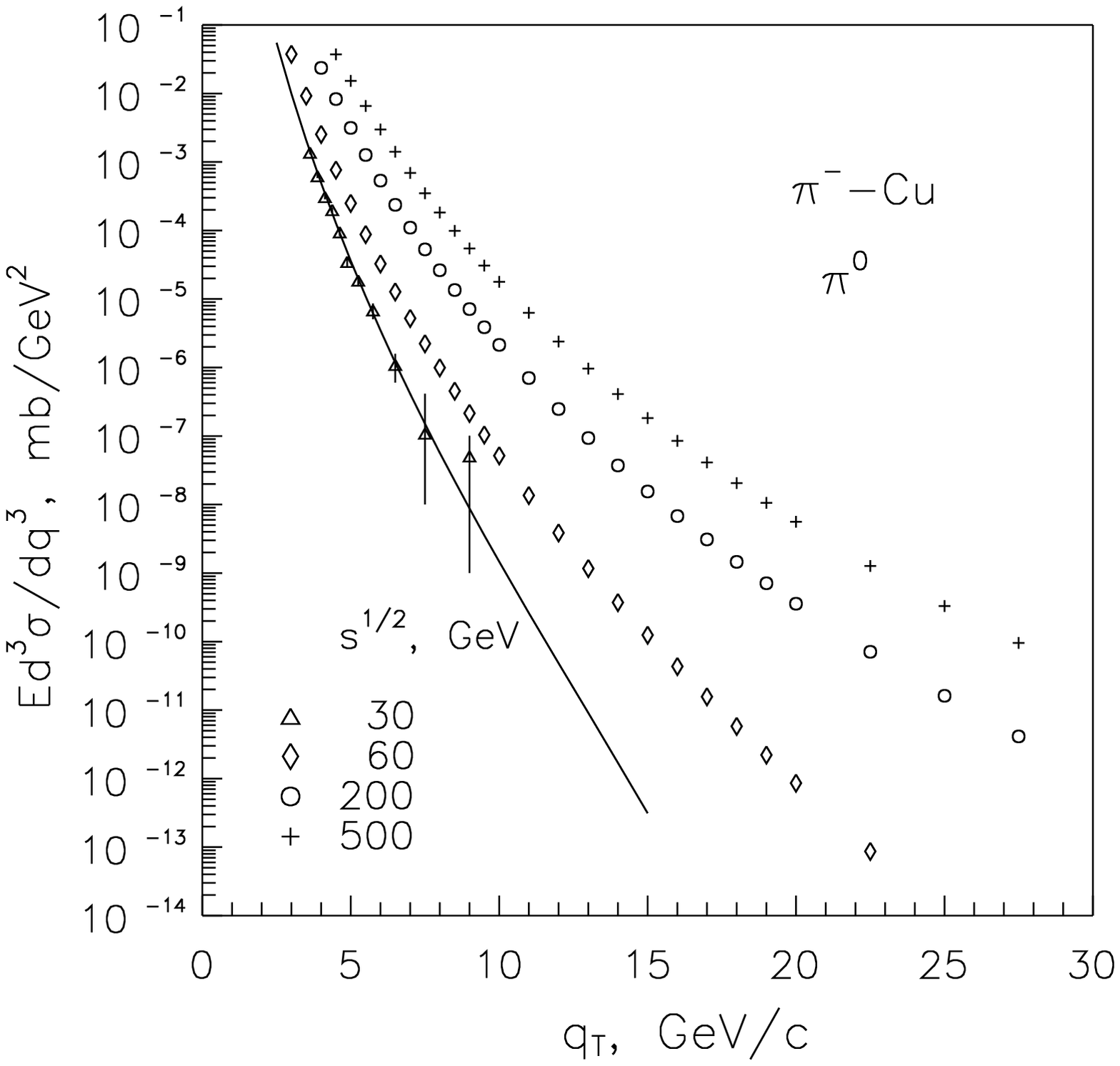}{}}
\vskip -1.cm
\hspace*{0.cm} a) \hspace*{8.cm} b)\\[0.5cm]
\end{center}

 {\bf Figure 6.}
  Dependence of the inclusive cross section of
 $\pi^0$-meson  production
 on transverse momentum $q_{T}$
 at $\theta_{cm}^{\pi N} \simeq 90^{0}$
 in
 $\pi^--Be$ (a),  $\pi^--Cu$ (b) and  $\pi^--Au$ (c) collisions.
 The calculated results are shown by
 points and solid lines $(\diamond - 60~GeV, \circ - 200~GeV,
 + - 500~GeV)$.
 Experimental data $(\triangle)$ are taken from \cite{alver93,E706}.

\end{document}